\documentclass[acmtog,nonacm]{acmart}

\usepackage{booktabs} 
\usepackage{appendix}

\usepackage[ruled]{algorithm2e} 

\SetAlFnt{\small}
\SetAlCapFnt{\small}
\SetAlCapNameFnt{\small}
\SetAlCapHSkip{0pt}

\acmJournal{TOG}

\definecolor{checkcolor}{RGB}{0,128,0}
\definecolor{headercolor}{RGB}{0,0,0}
\definecolor{lightgray}{RGB}{224,224,224}

\begin{document}


\title{Zero-Shot Human-Object Interaction Synthesis with Multimodal Priors}


\author{Yuke Lou}
\email{louyuke@connect.hku.hk}
\affiliation{%
  \institution{The University of Hong Kong}
  \country{China}
}
\authornote{equal contribution}

\author{Yiming Wang}
\email{wangyim@ethz.ch}
\affiliation{%
  \institution{ETH Zurich}
  \country{Switzerland}
}
\authornotemark[1]

\author{Zhen Wu}
\email{zhenwu@stanford.edu}
\affiliation{%
  \institution{Stanford University}
  \country{United States of America}
}

\author{Rui Zhao}
\email{reyzhao@tencent.com}
\affiliation{%
  \institution{Tencent}
  \country{China}
}

\author{Wenjia Wang}
\email{wwj2022@connect.hku.hk}
\affiliation{%
  \institution{The University of Hong Kong}
  \country{China}
}

\author{Mingyi Shi}
\email{irubbly@gmail.com}
\affiliation{%
  \institution{The University of Hong Kong}
  \country{China}
}

\author{Taku Komura}
\email{louyuke@connect.hku.hk}
\affiliation{%
  \institution{The University of Hong Kong}
  \country{China}
}
\authornote{corresponding author}

\authorsaddresses{} 

\begin{abstract}
Human-object interaction (HOI) synthesis is important for various applications, ranging from virtual reality to robotics. 
However, acquiring 3D HOI data is challenging due to its complexity and high cost, limiting existing methods to the narrow diversity of object types and interaction patterns in training datasets.
This paper proposes a novel zero-shot HOI synthesis framework without relying on end-to-end training on currently limited 3D HOI datasets. The core idea of our method lies in leveraging extensive HOI knowledge from pre-trained Multimodal Models. Given a text description, our system first obtains temporally consistent 2D HOI image sequences using image or video generation models, which are then uplifted to 3D HOI milestones of human and object poses. We employ pre-trained human pose estimation models to extract human poses and introduce a generalizable category-level 6-DoF estimation method to obtain the object poses from 2D HOI images. Our estimation method is adaptive to various object templates obtained from text-to-3D models or online retrieval.
A physics-based tracking of the 3D HOI kinematic milestone is further applied to refine both body motions and object poses, yielding more physically plausible HOI generation results. The experimental results demonstrate that our method is capable of generating open-vocabulary HOIs with physical realism and semantic diversity. 
Project Page: \href{https://thorin666.github.io/projects/ZeroHOI}{https://thorin666.github.io/projects/ZeroHOI}.
\end{abstract}



\begin{teaserfigure}
  \includegraphics[width=\textwidth]{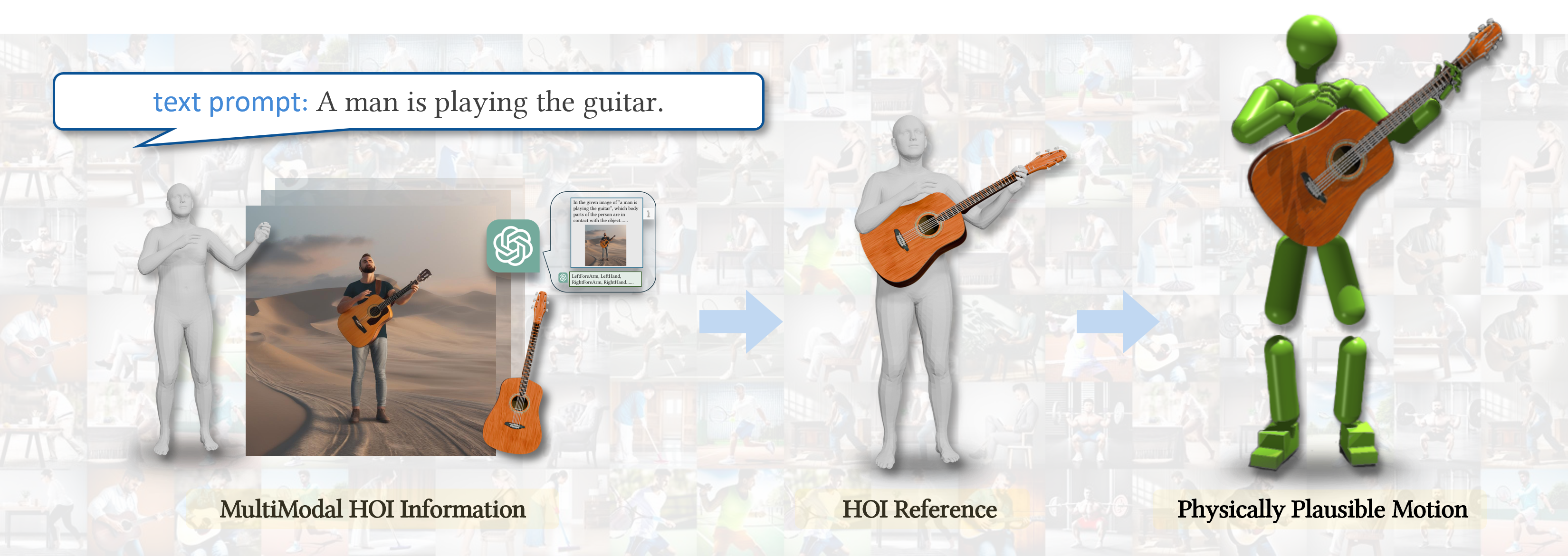}
  \caption{We present a system capable of generating human-object interactions without relying on 3D HOI data while generalizing to unseen objects.}
  \label{fig:teaser}
\end{teaserfigure}

\maketitle

\section{Introduction}
\label{sec:introduction}
With the advancements in diffusion models, recent text-to-motion generation frameworks~\cite{guo2023momask, jiang2024motiongpt} trained end-to-end on 3D motion datasets~\cite{AMASS} have demonstrated the ability to synthesize diverse motion sequences. However, these models face challenges in generating realistic human-object interaction (HOI) sequences due to the lack of explicit human-object interaction modeling. Furthermore, the limited availability of 3D HOI datasets further constrains the end-to-end training of HOI generation~\cite{karunratanakul2023gmd, karunratanakul2023guided}, limiting their ability to support a diverse range of object types and interaction patterns.

Compared to the cost and challenges of acquiring 3D data, especially the 3D HOI datasets, 2D images, videos, and text data are far more abundant and accessible.
Inspired by methods like DreamFusion~\cite{poole2022dreamfusion}, which leverage 2D diffusion models to generate 3D structures from textual descriptions, we explore adapting similar techniques for 3D HOI generation.
In particular, the development of ControlNet~\cite{zhang2023adding} has greatly improved the controllability of 2D diffusion generation, allowing us to specify human poses and generate corresponding 2D HOI content guided by textual descriptions.

In this paper, we present a novel optimization-based framework for zero-shot HOI generation using pre-trained multimodal models.
This framework operates without the need for end-to-end training on currently limited 3D HOI datasets, leveraging the extensive HOI information in Large Multimodal Models to facilitate the handling of a diverse range of object types and motion patterns.
Additionally, we integrate a physics-based simulator to refine the generated HOIs to be physically plausible.
Given a text description, our method is capable of simultaneously generating the corresponding motions for both the human and the object.

Our system begins by extracting existing 2D human-object interaction (HOI) priors embedded within state-of-the-art image and video generation models, which are specifically tailored to produce temporally consistent 2D HOI image sequences. 
The extracted 2D HOI knowledge is subsequently leveraged to uplift to 3D HOI milestones of human poses and object poses. We use pre-trained human pose estimation models to obtain the 3D human poses. 
Given arbitrary object templates generated by text-to-3D models or obtained from online sources using the input text prompt,
we propose a generalizable category-level object 6-DoF estimation method to extract the object poses from the generated 2D HOI images. This method employs a two-stage optimization process to address potential geometric and appearance variations between the input object template and the generated 2D HOI images: an initial coarse estimation obtained by solving the Perspective-n-Point (PnP) problem using semantic correspondences, followed by a refinement stage employing differentiable rendering. 
We then conduct physics-based tracking\cite{peng2018deepmimic, peng2021amp} of the synthesized 3D HOI milestones of body motion and object pose within a physics simulation environment, resulting in a physics-plausible animation that accurately depicts the hands interacting with the object.

Compared to other HOI generation methods trained on 3D HOI data, which are typically constrained by the object types and HOI patterns observed in currently limited 3D HOI datasets, our zero-shot generation framework leverages extensive 2D, 3D and textual HOI information in large multi-modal Models trained on much larger scale datasets. Building on this advantage, our approach is applicable to a more diverse range of objects and capable of generating a broader spectrum of HOIs. By incorporating refinement within a physics simulation environment, we further enhance the physical realism of the generated HOI. Comparative evaluations against baseline methods demonstrate the superior capacity of our approach to produce more realistic and diverse HOI outcomes.
Furthermore, our system is highly versatile, capable of not only generating HOIs but also augmenting existing ground truth human motions with objects, reconstructing HOIs from video footage, and can be further utilized for automatic 3D HOI dataset generation.

In summary, our main contributions in this paper can be summarized as follows:
\begin{itemize}
    \item We introduce an innovative zero-shot human-object interaction (HOI) generation framework that leverages extensive HOI knowledge from pre-trained multi-modal models. 
    \item We propose a generalizable category-level object 6-DoF estimation method that effectively adapts to various object templates and synthesizes 2D HOI images from text inputs.
    \item We integrate the proposed zero-shot HOI generation method with a physics-based tracking strategy, enabling our method to achieve both diverse and physically
    realistic HOI generation.
\end{itemize}
\section{Related Work}
\label{sec:related_work}
In this section, we discuss prior research in related fields. We first review methods for text-to-motion generation and human-object interaction (HOI) synthesis. Subsequently, we introduce works on physics-based animation. Finally, we discuss the use of priors in 3D generation methods.

\subsection{Text2Motion Synthesis}
As the field of motion synthesis continues to advance, researchers are exploring the use of various modalities of information as conditions to enhance controllability. Among these modalities, text has become one of the most widely used, leading to a growing interest in text-guided motion synthesis.
The availability of large-scale motion capture datasets such as AMASS~\cite{AMASS}, BABEL~\cite{punnakkal2021babel}, and HumanML3D~\cite{Guo_2022_CVPR} has paved the way for new developments in motion synthesis driven by actions and text~\cite{petrovich2021action, Guo_2022_CVPR, petrovich2022temos, tevet2022motionclip}. It has been shown that using VAEs is an effective approach for creating varied human motions from text descriptions~\cite{Guo_2022_CVPR, guo2022tm2t}. More recently, diffusion models have shown promise in this area~\cite{chen2023executing, barquero2023belfusion, huang2023diffusion, raab2023single, shafir2023human, yuan2023physdiff, zhang2023tedi, li2023object, shi2023controllable}, leading to substantial research on generating motions from text with precise control~\cite{tevet2022human, dabral2022mofusion, zhang2022motiondiffuse, karunratanakul2023gmd, guo2023momask}.
In this work, we also take language descriptions as input to guide our 3D human-object interaction generation. Instead of synthesizing human motion alone, we generate both object motion and human motion conditioned on the text.

\subsection{Human-Object Interaction}
\label{subsec:hoi}
Humans interact with objects constantly, making the generation of human-object interactions a crucial aspect of character animation.
Consequently, various approaches have been proposed to generate and reconstruct HOIs.
Some studies have focused on reconstructing HOIs from video~\cite{li2019estimating, ehsani2020use, ye2023diffusion}.
Others limit their scope to interactions between humans and static scenes~\cite{zhang2020generating, yi2022human, hassan2021stochastic, wang2024sims}.
For HOI generation, different settings have been explored.
For instance, \cite{li2024task, ye2023affordance, zhou2022toch} focus exclusively on hand-object interactions.
Given the object,~\cite{li2023object} predicts the corresponding human motion.
Studies such as~\cite{ghosh2023imos, li2024task} target fundamental HOIs, such as moving objects.
Additionally, diffusion models have recently been employed to generate high-quality HOIs~\cite{xu2023interdiff, peng2023hoi}.

However, compared to the increasingly mature technology of human motion capture, capturing human-object interactions remains significantly more challenging and currently lacks accessible, low-cost solutions.
As a result, existing datasets for human-object interactions~\cite{bhatnagar2022behave, wan2022learn, mandery2015kit, taheri2020grab} feature a limited variety of objects and constrained interaction patterns between humans and objects.
CHOIS~\cite{li2023controllable} demonstrates the ability to generate object and human motions simultaneously from language descriptions, but it still relies on supplementary information such as waypoints.
Similarly, InterDiff~\cite{xu2023interdiff} exhibits some level of generalization but is restricted to handling objects with similar shapes.

\subsection{Physics-based Animation}
\label{subsec:physcis}

Compared to kinematic methods, physics-based animation~\cite{peng2018deepmimic, peng2021amp, peng2022ase, luo2023perpetual} incorporates physical constraints to control agent movements within a simulated environment, effectively addressing issues such as sliding and penetration.
Because physics-based methods can produce physically realistic results, they have been widely adopted for human-object interaction (HOI) synthesis.
Examples include interacting with scenes~\cite{hassan2023synthesizing, pan2023synthesizing, xiao2023unified}, playing basketball~\cite{liu2018learning, wang2023physhoi}, playing soccer~\cite{hong2019physics, xie2022learning}, playing tennis~\cite{yuan2023learning}, catching and carrying~\cite{merel2020catch}, using chopsticks~\cite{yang2022learning}, and multi-character interactions~\cite{zhang2023simulation}.
However, most of these works are tailored to specific object types, and only a few frameworks are designed to be universal and task-agnostic.

\subsection{Utilizing 2D and Language Priors}
\label{subsec:zeroshot}

A significant challenge in 3D-related tasks is the difficulty of acquiring 3D data compared to 2D images or text, resulting in generally smaller datasets.
To address this limitation, an increasing number of studies exploit external knowledge to facilitate 3D content generation.
For example, pretrained 2D text-to-image diffusion models have been successfully used in text-to-3D synthesis to alleviate the scarcity of labeled 3D data~\cite{poole2022dreamfusion, wang2023score}.
For 3D human motion, 2D images have also been utilized to reconstruct dynamic interactions.
For instance, \cite{muller2023generative} learn a prior for reconstructing 3D social interactions.
\cite{li2023genzi} and \cite{coma} estimate human presence based on the surrounding environment and objects in 2D images. In contrast, our approach builds on human poses to infer objects using pre-trained 2D diffusion models, providing a more intuitive and accurate way to generate plausible 2D HOI images.
Additionally, large language models (LLMs) have been explored to facilitate HOI tasks. ~\cite{wang2022reconstructing} utilize LLMs to infer contact points between the human body and object. It focuses on estimating human and object poses from in-the-wild videos where the object template is provided. In contrast, our system uses object template from Text-to-3D models and estimates its pose from 2D generated images and videos, accounting for potential geometric and appearance variations between the input object templates and the 2D generated HOI data.
InterDreamer~\cite{xu2024interdreamer} performs zero-shot HOI generation by leveraging LLMs for text-based analysis. However, it does not utilize extensive 2D HOI data and relies solely on text, resulting in suboptimal results.

\begin{figure*}[t]
    \centering
    \includegraphics[width=\textwidth]{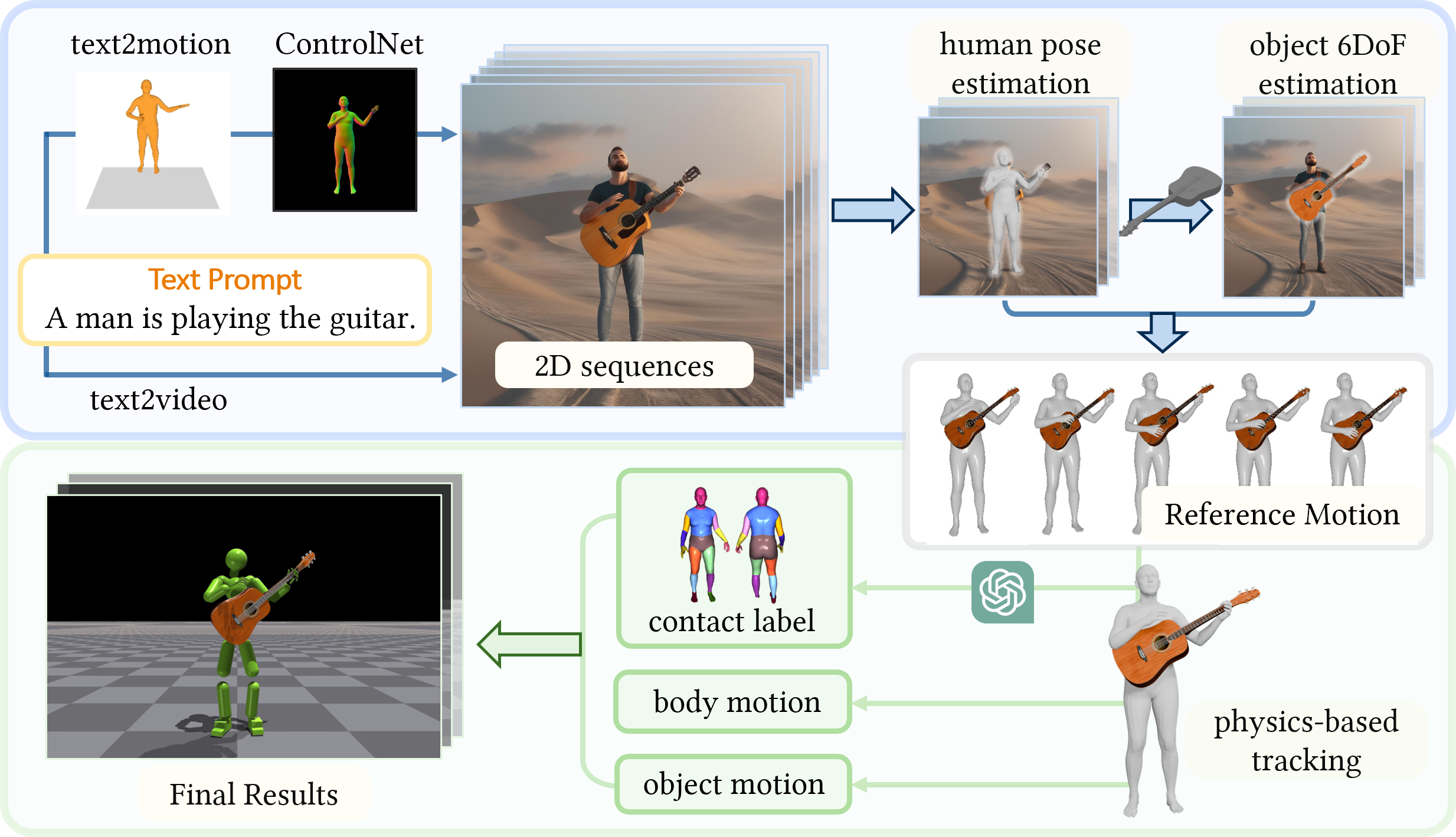}
    \caption{
    Our system is composed of two core components: 
    (a) a zero-shot HOI generation pipeline that leverages the generative capabilities of pre-trained multimodal models to obtain rough 3D interaction between humans and objects from text input;
    (b) a physics-based tracking strategy applied to the HOI generated in part (a) to produce physically plausible animations.
    }
    \Description{}
    \label{fig:system_overview}
\end{figure*}
\section{System Overview}
\label{sec:system_overview}

Our system takes textual descriptions as input to generate diverse and realistic human-object interactions (HOIs) in a zero-shot manner. 
As illustrated in Fig.~\ref{fig:system_overview}, our system can be divided into two structural parts: 
\textbf{(a)} the first part (Sec~\ref{sec:hoi_reconstruction}) utilizes the generative capability of existing large multi-modal models, extending their knowledge to obtain rough 3D interaction between humans and objects from text input; 
\textbf{(b)} the second part (Sec~\ref{sec:motion_tracking}) utilizes physics-based tracking to generate physically realistic and contact-rich animations of human-object interactions given the coarse 3D interaction obtained from part (a).

Our system first generates temporally consistent 2D HOI image sequences using image generation models or video generation models, with the image generation models enhanced by conditioning the generation process on human poses derived from a text-to-motion model. We then uplift the obtained 2D HOI knowledge to 3D HOI milestones of human poses and object poses. We use pre-trained human pose estimation models to obtain human poses from the 2D HOI images. 
Considering that the object template derived from text-to-3D models or the Internet can differ in appearance and geometric details from the generated 2D HOI images, we develop a generalizable category-level object 6-DoF estimation method to adapt to various object identities in 2D HOI images. 
The generated human and object motions are then used as reference motions for the physics-based tracking component.

In the second component, RL training is conducted in IsaacGym~\cite{makoviychuk2021isaac} to develop a control policy that mimics the reference motion. Simultaneously, LLMs~\cite{gpt4o,liu2023llava} are employed to generate contact labels for human-object interactions, which are integrated into the reward function to optimize training. This process will result in a final physically realistic motion that matches the interaction between the human and objects. We will elaborate on these two parts in Sec.~\ref{sec:hoi_reconstruction} and Sec.~\ref{sec:motion_tracking}, respectively.

\section{Zero-shot HOI Generation}
\label{sec:hoi_reconstruction}

Given a text description of a human interacting with a specific object, such as \textit{"A man is playing the guitar"}, our goal is to generate a $N$-frame-long sequence consisting of full-body human motion $\{h_{i}\}_{i=1}^N$ and object 6-DoF poses $\{o{i}\}_{i=1}^N$ in a zero-shot manner. 

To achieve this challenging goal without relying on training models with 3D HOI data, our key insight is to leverage the widespread 2D human-object interaction knowledge in 2D generative models pre-trained on large-scale 2D datasets. As shown in Fig.~\ref{fig:system_overview} (a), our system first obtains temporally consistent 2D HOI milestones from image or video generation models (Sec ~\ref{subsec:2dhoi_gen}). It then extracts human poses using pre-trained human pose estimation models (Sec ~\ref{subsec:human_motion_gen}) and estimates object poses through a generalizable category-level object 6-DoF estimation method (Sec ~\ref{subsec:6dof_estimation}). 

\subsection{2D HOI Milestones Generation}
\label{subsec:2dhoi_gen}

Advancements in 2D diffusion models~\cite{rombach2022high} trained on large-scale 2D datasets enable the generation of high-quality 2D HOI images and videos, which can serve as a sufficient source of information for generating 3D HOI data. 
While video generation models excel at producing temporally consistent 2D HOI image sequences, the relatively larger size of existing image datasets allows image generation models to achieve better control over aspects such as camera view and produce higher-quality results. 
To fully utilize the available 2D HOI sources, our system is designed to effectively leverage 2D HOI priors from both image and video generation models. Below, we detail how our system incorporates and utilizes these models accordingly.

\subsubsection{Generative 2D HOI Images}
\label{subsubsec:gen_2d_image}
The key challenge in leveraging the image diffusion model to produce 3D HOI sequences lies in generating a series of temporally consistent 2D HOI images. To solve this problem, we use ControlNet~\cite{zhang2023adding} to condition the 2D HOI diffusion generation on 2D human motion sequences. Specifically, we first use pre-trained Text-to-Motion models to synthesize initial human motion sequences from the text prompt, and then uniformly extract keyframe poses as the 2D diffusion condition.  
Instead of using the original ControlNet's~\cite{zhang2023adding} mode that uses a skeleton's 2D keypoints as a condition, we use normal images rendered from a human mesh as diffusion condition following~\cite{ge2024humanwild}, which provides more accurate and detailed control. 
The generated 2D HOI images will serve as milestone inputs for generating human motion and object poses in subsequent sections.

In contrast to 2D HOI diffusion that is conditioned on rendered object images\cite{li2023genzi, coma} which typically requires accurate initialization of accurate object poses, our method adopts a human-centric strategy, generating objects based on the human body. As shown in Fig.~\ref{fig:coma2d}, this approach enables 2D generation models to more easily produce plausible 2D HOI images by leveraging accurate human poses from Text-to-Motion models, rather than relying on heuristically initialized object poses.

\subsubsection{Generative 2D HOI Videos}
Current video generation models, such as Kling~\cite{KLING} and SORA~\cite{Sora}, demonstrate significant potential in generating high-quality videos based on text or image inputs. In our study, we explored two setups for obtaining 2D HOI videos using video generation models: one with text input alone and the other combining text input with a start-frame image obtained from the generative 2D HOI image pipeline. The primary advantage of using an additional start-frame image as a condition is that it allows control over the rendered camera view, ensuring the HOI's region of interest is prominently displayed.
We then uniformly sample keyframe images from the generated video as 2D HOI image milestones.

\subsection{Human Motion Generation}
\label{subsec:human_motion_gen}

After obtaining the corresponding 2D HOI milestones given the text input, we apply pre-trained human pose estimation models to obtain the human poses. 
We use TRAM~\cite{wang2024tram} to estimate the global human trajectory and human motion jointly from generated 2D HOI videos. The 3D human milestone poses are uniformly sampled from the extracted human motion. 

Image generation models conditioned on continuous motion input can produce semantically consistent 2D HOI image sequences but often exhibit temporally discontinuous details, which limits the performance of video pose estimation models, such as TRAM. To address this issue, we employ SMPLer-X~\cite{cai2024smplerx} to estimate the local human motion from each frame of the generated 2D HOI images, replacing the local motion generated by the text-to-motion model while preserving its global human trajectory.

\begin{figure}[h]
    \centering
    \includegraphics[width=\linewidth]{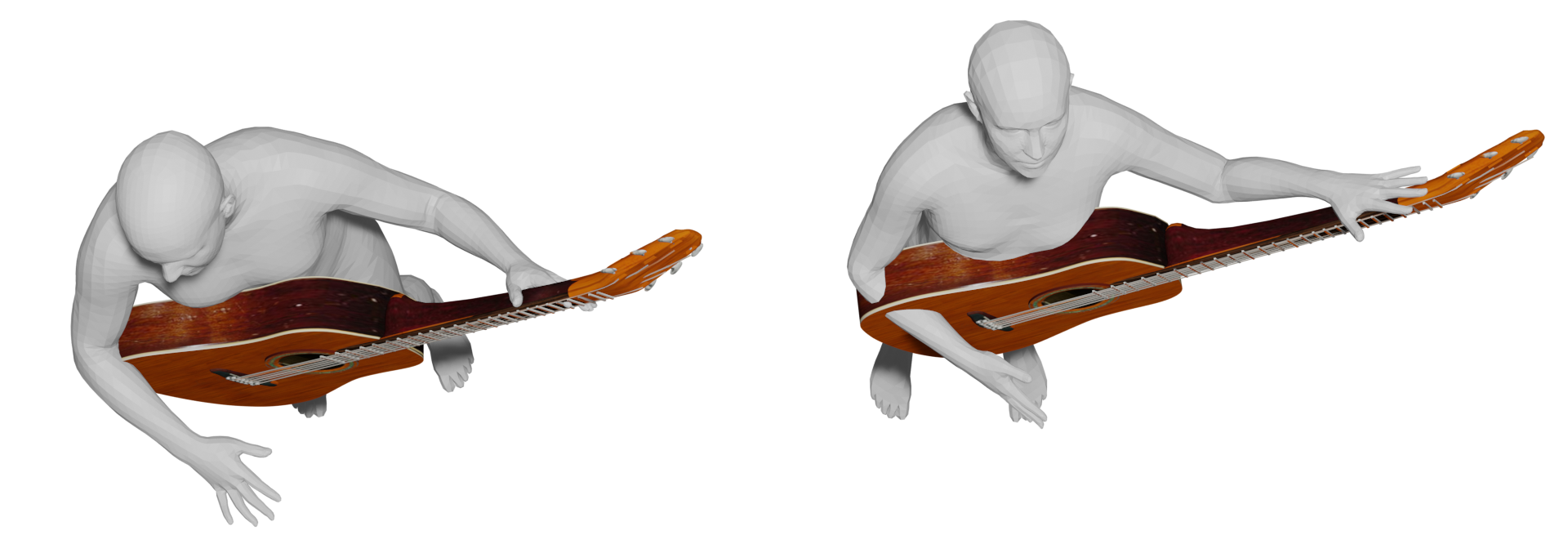}
    \caption{\textbf{Left: Rectified Pose. Right: Initial Pose.} Human motions generated by Text-to-Motion models may lack spatial awareness of objects, which limits the effectiveness of subsequent human-object interaction optimization. For instance, given the prompt \textit{"A man is playing the guitar"}, the generated human body motion fails to provide sufficient space for a plausible guitar placement. 
    Additional examples illustrating the benefits of motion rectification are provided in Fig.~\ref{fig:rec}.}
    
    \Description{}
    \label{fig:rectify}
\end{figure}

We do not directly use the full human motion generated by Text-to-Motion models because it often mismatches with the object. These issues arise from the model's training on datasets that include only human motion without object context. Therefore, we use the aforementioned human estimation pipeline to rectify the human motion using the generated 2D HOI images, incorporating human poses that account for object interaction. The effect of motion rectifying is presented in Fig.~\ref{fig:rectify}.

\subsection{Category-level Object 6-DoF Estimation }
\label{subsec:6dof_estimation}
After obtaining the 3D human motions, our next objective is to estimate the corresponding object 6-DoF poses in the 2D HOI milestones. 
Given an object template that can be obtained either retrieving from a large object dataset~\cite{objaverseXL} or current text-to-3D models~\cite{xu2024grm, tang2024lgm, wei2024meshlrm}, 
the primary challenge in this specific object 6-DoF estimation task lies in the potential geometric variations between the input object template and the generated 2D HOI images. To tackle this challenging task, we propose a novel two-stage optimization pipeline designed to maximize the use of category correspondence information. In contrast, most existing object pose estimation methods~\cite{bhatnagar2022behave, xie2022chore, xie2023visibility, wang2022reconstructing} rely on object templates that precisely match those in the input images.

In the first stage (Sec~\ref{subsubsec:semantic_correspondance}), we use semantic correspondence extracted from a pretrained 2D vision model Dinov2~\cite{oquab2023dinov2} to get the object 6-DoF approximation by solving the Perspective-n-Point (PnP) problem.
In the next stage (Sec~\ref{subsubsec:diff_rendering}, Sec~\ref{subsubsec:hoi_optimization}), we refine the object 6-DoF pose using the PyTorch3D~\cite{ravi2020pytorch3d} differentiable renderer
that optimizes both silhouette and depth, integrated with 3D human priors and contact labels.

\subsubsection{Semantic Correspondence}
\label{subsubsec:semantic_correspondance}
Recent self-supervised learning methods~\cite{oquab2023dinov2} and image diffusion models~\cite{tang2023emergent} have shown great potential in extracting general-purpose visual features, which are especially useful for building image correspondences. Inspired by these works, we use the extracted visual features from Dinov2~\cite{oquab2023dinov2} to build dense semantic correspondences for the object template and synthesized 2D HOI images. 

In order to get the visual feature descriptor for the object template, we first need to render the object template using a camera viewpoint that reflects the entire object as much as possible. Inspired Gen6d~\cite{liu2022gen6d}, we employ a viewpoint selector that renders the object from 24 distinct viewpoints and identifies the viewpoint with the highest similarity to the 2D HOI image. 
The similarity is measured as the mean Euclidean distance between the visual feature vectors of the rendered object image and the 2D HOI image, both extracted using Dinov2.

We apply a bidirectional matching algorithm to the visual descriptors extracted from the rendered image and the HOI image, using the Euclidean distance of DinoV2 features as the similarity metric. 
A homogeneous transformation between the matched descriptors is then estimated with the Random Sample Consensus (RANSAC) algorithm, effectively filtering out outliers and selecting a subset of reliable correspondences. Finally, we solve the Perspective-n-Point (PnP) problem using the inlier correspondences to compute the 6-DoF pose to align the object with the 2D HOI images.

\subsubsection{Differentiable Rendering}
\label{subsubsec:diff_rendering}
We further refine the object pose by leveraging the object's silhouette and depth information present in the 2D HOI image. Additionally, we incorporate existing 3D human prior to enhance the accuracy of our object pose estimation.

We use the PyTorch3D~\cite{ravi2020pytorch3d} differentiable renderer to render the human-object silhouette $\mathcal{S}$ and object silhouette $\mathcal{S}_o$. We use rembg~\cite{Rembg} to extract the foreground human-object mask $\hat{\mathcal{S}}$ from the 2D HOI image and use SegmentAnything~\cite{kirillov2023segment} to extract the object mask $\hat{\mathcal{S}_o}$. 
To enhance the accuracy of the object mask, we utilize the matched object descriptors obtained in Sec~\ref{subsubsec:semantic_correspondance} as input labels for object segmentation. The overall silhouette loss is represented as:

\begin{equation}\label{loss_silhouette}
L_{sil} = \left| \mathcal{S} - \hat{\mathcal{S}} \right| +  \lambda_{object} \left| \mathcal{S}_o - \hat{\mathcal{S}_o} \right|
\end{equation}
where $\lambda_{object}$ is the mask confidence output from SegmentAnything.

We also use the estimated relative depth $\hat{\mathcal{D}}$ obtained from a monocular depth estimation model~\cite{yang2024depth} to supervise the rendered depth $\mathcal{D}$. Following \cite{li2020nsff, ranftl2020towards}, we use a robust scale-shift invariant loss function for the depth supervision. This loss function involves a robust estimator $E^*$, which normalizes the depths to have zero translation and unit scale:

\begin{equation}\label{loss_depth_estimator}
E^*(\mathcal{D}) = \frac{\mathcal{D} - \text{median}(\mathcal{D})}{\text{mean}(\left |\mathcal{D} - \text{median}(\mathcal{D})\right|)}.
\end{equation}
The overall relative depth loss is represented as:

\begin{equation}\label{loss_depth}
L_{depth}^{rel} = \left| E^*(\mathcal{D}) - E^*(\hat{\mathcal{D}}) \right| + \lambda_{object} \left| E^*(\mathcal{D}_o) - E^*(\hat{\mathcal{D}}_o) \right|
\end{equation}
where $\mathcal{D}_o$ is the object depth obtained using the object mask $\hat{\mathcal{S}_o}$.

While the aforementioned depth prior provides only relative depth information, we further incorporate metric depth priors from the 3D human model to ensure that the object does not appear too distant from the human:

\begin{equation}\label{loss_human_depth}
    L_{depth}^{abs} = \left | \text{mean}(\mathcal{D}_o) - \mathcal{D}_h \right |
\end{equation}
where $\mathcal{D}_h$ is the mean depth of the human body.

\subsubsection{Human-Object Interaction Optimization}
\label{subsubsec:hoi_optimization}
Unlike conventional 6-DoF estimation tasks~\cite{zhang2020perceiving} that focus solely on the object, our task places significant emphasis on the interaction between the human body and the object.
Therefore, we incorporate a set of human-object interaction loss functions that leverage human body mesh information. This approach further enhances the precision of object pose estimation, particularly in the context of HOI, thereby enabling more effective learning in the physics-based tracking stage.

\paragraph{Hand Contact Loss}
Considering the prevalence of hand interactions in human-object interactions (HOIs), we propose a targeted loss function specifically for the hands to enhance performance in these scenarios. 
We utilize LLMs~\cite{dubey2024llama, gpt4o} to derive hand contact labels $w_{hand}$ from textual descriptions. These labels indicate whether the left or right hand remains in contact with the object during the whole interaction. The strategy of obtaining contact labels will be detailed in Section~\ref{sec:motion_tracking}.

Our objective is to minimize the distance between the object and the palm of the hand. The contact loss is defined as follows:

\begin{equation}\label{loss_contact2}
L_{contact} = w_{hand} \sum_{j \in V_{palm}} \mathcal{H}(\theta - d(\mathbf{p}_j, \mathbf{M}_{object})) |d(\mathbf{p}_j, \mathbf{M}_{object})|,
\end{equation}
where $w_{hand}$ is a binary flag indicating whether the hand is in contact with the object during the whole motion, \( V_{palm} \) represents the set of vertices on the human mesh's palm, \( \mathbf{p}_j \) is the position of the \( j \)-th palm vertex, and \( \mathbf{M}_{object} \) is the mesh of the object. The distance function \( d(\mathbf{p}_j, \mathbf{M}_{object}) \) calculates the distance from the palm vertex to the object mesh, and \( \theta \) is a predefined threshold for valid contact regions. The usage of the Heaviside step function \( \mathcal{H} \) ensures that the loss is only applied when the palm vertices are within a certain proximity to the object, thus encouraging a realistic interaction where the hand appears to be in contact with the object.

\paragraph{Penetration Loss}
Considering that penetration issues can significantly reduce the realism of the generated results and may lead to undesirable consequences in the physics engine, we design a loss function to avoid penetration, which is defined as follows.

\begin{equation}\label{loss_penetration}
L_{penetration} = \sum_{i \in V_{object}} \max(0, -d(\mathbf{p}_i, \mathbf{M}_{human})),
\end{equation}
where \( V_{object} \) denotes the set of vertices on the object, and \( \mathbf{p}_i \) is the position of the \( i \)-th vertex on the object. The human mesh is represented by \( \mathbf{M}_{human} \), and \( d(\mathbf{p}_i, \mathbf{M}_{human}) \) is the signed distance function from vertex \( i \) to the human mesh, which is negative when the vertex is inside the mesh. The \( \max \) function ensures that only negative distances, indicating penetration, contribute to the loss, by adding the absolute value of such distances.

Following the process outlined above, we have acquired the 3D keyframe HOI milestones of human motion and object poses, which  will serve as reference HOIs for physical tracking in the next section. 
To facilitate more effective tracking within the physical simulation, we further convert sparse rewards from milestone motion into dense rewards by interpolating the human and object poses into smooth, continuous motion within keyframe milestones. 

\section{Physics Based HOI Refinement}
\label{sec:motion_tracking}

Despite various optimizations, the generated human-object interactions from the aforementioned pipeline still lack physical realism. To address this, we incorporate an imitation learning policy within a reinforcement learning framework to track reference motions in a simulated environment. 
In the following context, we refer to the obtained 3D HOI milestone as the reference motion.
\begin{figure}[h]
    \centering
    \includegraphics[width=\linewidth]{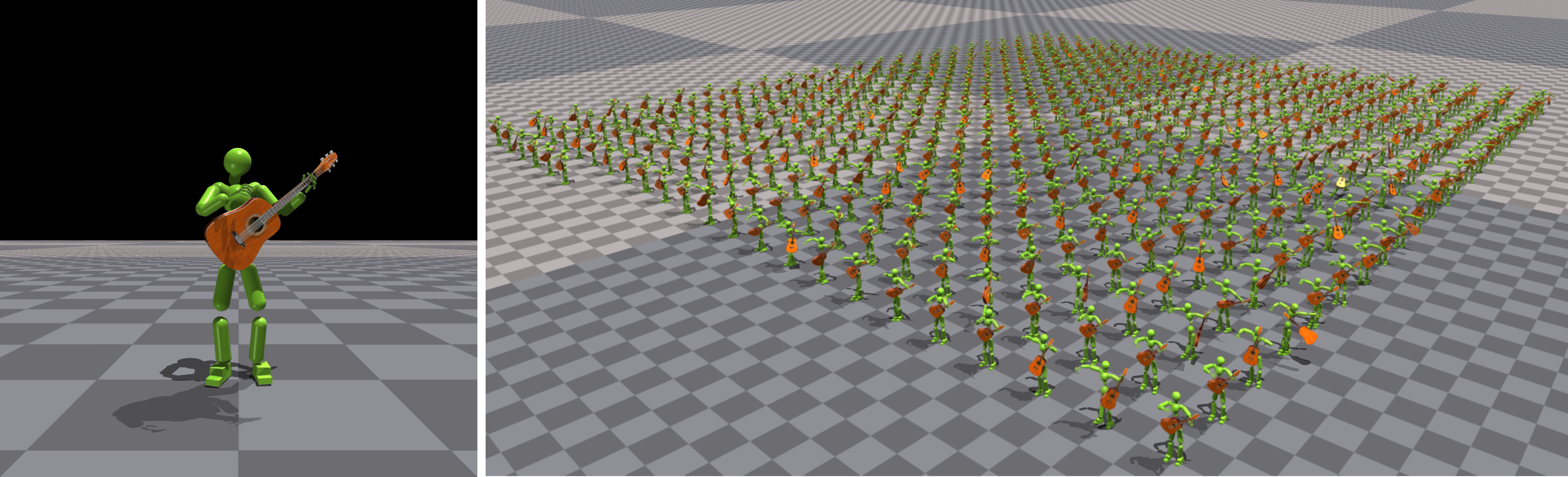}
    \caption{We train a control policy in Isaac Gym to mimic the reference motion.}
    \Description{}
    \label{fig:physics}
\end{figure}

Building on DeepMimic~\cite{peng2018deepmimic}, we conduct a physics-based tracking of the generated 3D HOI milestone, including the human and object poses.
Compared to the original DeepMimic, we introduce two key advancements. First, we constrain object motion to align with the reference motion, ensuring realistic body and hand movements that effectively fulfill the HOI tasks. Second, we integrate LLMs to generate high-level contact plans between the body and objects. This serves as an additional reward function, enabling more precise body-object contact and further enhancing the realism of the interactions.

Our method employs reinforcement learning, in which the agent interacts with its environment guided by a policy designed to maximize rewards. At each timestep \( t \), the agent receives the system states \( s_t \) as inputs and generates an action \( a_t \) by sampling from the policy distribution \( \pi(a_t | s_t) \). Utilizing the physics simulator function \( f(s_{t+1} | a_t, s_t) \), the chosen action \( a_t \) leads to a new state \( s_{t+1} \). Subsequently, a reward \( r_t = r(s_t, a_t, s_{t+1}) \) is computed. The objective is to develop a policy that maximizes the expected return 
\[
R(\pi) = \mathbb{E}_{p_\pi(\tau)} \left[ \sum_{t=0}^{T-1} \gamma^t r_t \right],
\]
where \( \tau = \{s_0, a_0, r_0, \ldots, s_{T-1}, a_{T-1}, r_{T-1}, s_T\} \) denotes the trajectory, and \( p_\pi(\tau) \) is the probability density function of the trajectory. Here, \( T \) represents the time horizon of a trajectory, and \( \gamma \), ranging from 0 to 1, is the discount factor.
We further discuss the state and reward function used in our policy in the following part of the section.

\subsection{HOI State Representation}
The state includes features describing the character's pose and the relative arrangement of objects in the scene. These features include the root position and rotation, root linear and angular velocity, local joint rotations, local joint velocities, positions of key joints (right hand, left hand, right foot, and left foot), object position and rotation, object linear and angular velocity, and hand contact force.
Please refer to the appendix for a detailed explanation of these variables and the methods employed to process them.

\subsection{Tracking Reward for Body and Object}
The reward function for the agent body and the object considers the difference between the states of the object and the agent in the simulation environment and the reference motion. For the agent, it is defined as follows:
\begin{itemize}
\item \textbf{Position Reward}: Encourages matching the position of key joints with the reference motion.
\item \textbf{Rotation Reward}: Aims to align joint rotations with the reference.
\item \textbf{Velocity Reward}: Compares actual linear velocities to reference values.
\item \textbf{Angular Velocity Reward}: Compares actual angular velocities to reference values.
\end{itemize}

The overall reward function can be expressed as:

\begin{align}
R_{body} &= \exp\left(\lambda_j \left[ \sum_e \left\| \hat{p}_t^{e} - p_t^{re} \right\|^2 \right] + \lambda_p \left[ \sum_j \left\| \hat{q}_t^{aj} \otimes \hat{q}_t^{rj} \right\|^2 \right] \right. \nonumber \\
&\quad \left. + \lambda_{v} \left[ \sum_j \left\| \hat{v}_t^{j} - v_t^{rj} \right\|^2 \right] + \lambda_{\omega} \left[ \sum_j \left\| \hat{\omega}_t^{j} - \omega_t^{rj} \right\|^2 \right] \right),
\end{align}
where \( \lambda_j \), \( \lambda_p \), \( \lambda_{v} \), and \( \lambda_{\omega} \) are weighting factors for the body position, pose, linear velocity, and angular velocity rewards, respectively, and the terms \( \hat{p}_t^{e} \), \( p_t^{re} \), \( \hat{q}_t^{aj} \), \( q_t^{rj} \), \( \hat{v}_t^{j} \), \( v_t^{rj} \), \( \hat{\omega}_t^{j} \), and \( \omega_t^{rj} \) denote the corresponding quantities in the simulation environment and reference.
For objects, we also calculate the differences in position, orientation, and velocity relative to their reference values as the reward function $R_{obj}$.

Unlike typical tracking problems, our task uses generated reference motion, which can introduce jitter due to its lower quality compared to motion capture data. To balance similarity to the reference and motion smoothness, we introduce two regularization terms to constrain the control policy and joint accelerations, reducing unnecessary movements and jitter for better results:
\begin{align}
R_{reg} &= \exp\left(\lambda_{\text{action}} \| a_t \| + \lambda_{\text{acc}} \sum_j \| v_t^j - v_{t-1}^j \| \right).
\end{align}
This formula represents a regularization term designed to improve motion smoothness and stability. The first term measures the magnitude of the control policy output, scaled by the coefficient \( \lambda_{\text{action}} \). The second term calculates the sum of velocity differences between consecutive timesteps for all joints, representing joint accelerations, scaled by the coefficient \( \lambda_{\text{acc}} \). This regularization penalizes large control outputs and rapid changes in joint accelerations, promoting smoother and more natural motion.

In total, we use the product of the aforementioned rewards as the imitation reward, ensuring that both the agent and the object are tracking their reference motions:
\begin{equation}
R_{imitate}  = R_{body} \cdot R_{obj}\cdot R_{reg}
\end{equation}

\subsection{Contact Reward from LLM}
\label{subsec:contact_labelling}
Tracking human-object interactions is significantly more challenging than modeling simple human motion, especially when object trajectories are reconstructed rather than given as ground truth. In such cases, interactions learned purely through imitation may deviate from the desired behavior. For instance, while the goal may be to hold an object in front of the chest, the learned interaction might instead involve clamping the object between the hand and chest. Therefore, contact information is essential to guide and refine the desired human-object interaction.

Recently, LLMs have demonstrated exceptional performance in various tasks, including capturing interactions between humans and objects. Providing HOI descriptions to LLMs helps identify body parts in continuous contact with objects during movement. These contact labels are then compared with simulation data to calculate a contact reward, bridging the gap between abstract descriptions and physical realism.
\begin{equation}\label{reward_contact}
R_{contact} = \exp \left( \lambda_{contact}\sum_{j} \left| \mathbb{I}(|F_j| < \text{threshold}) - L_j \right| \right)
\end{equation}
where $\mathbb{I}$ is the indicator function that is 1 if the force magnitude $|F_i|$ is below the threshold, indicating no contact, and 0 otherwise. $|F_j|$ represents the force exerted on body part $j$, obtained through Isaac Gym. More specific settings can be found in the appendix.

\begin{figure*}[h]
    \centering
    \includegraphics[width=\textwidth]{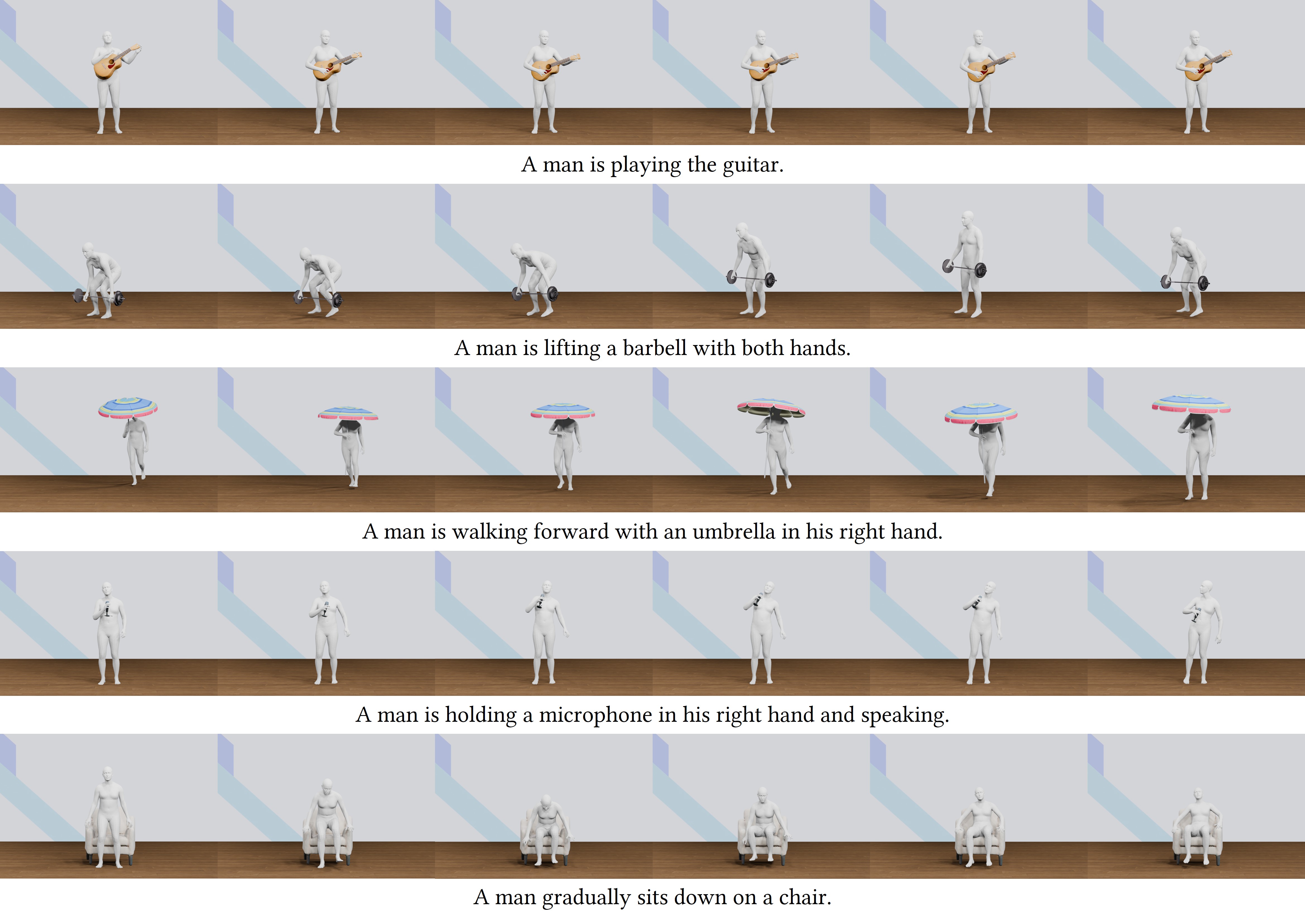}
    \caption{Zero-shot human-object interaction results generated by our system, using generative 2D Image pipeline.}
    \Description{}
    \label{fig:results}
\end{figure*}

\section{Results}
\label{sec:results}
In this section, we first provide the implementation details for our system setup (Sec~\ref{subsec:system_setup}), followed by qualitative and quantitative comparisons of our zero-shot HOI synthesis results with baseline methods (Sec~\ref{subsec:zeroshot synthesis}). 
Next, we analyze the effectiveness of our system's key components (Sec~\ref{subsec:system_analysis}). 
Finally, we analyze the system's success rate and show examples of failure cases (Sec~\ref{subsec:failure}).

\subsection{System Setup}
\label{subsec:system_setup}

Our system supports two approaches for obtaining 2D HOI priors from pre-trained models: Text-to-Image models and Text-to-Video models. We use KLING~\cite{KLING} as our video generation model, and employ Tram~\cite{wang2024tram} to estimate the human motion from the videos. For the Text-to-Image pipeline, we first use Text-to-Motion models MotionGPT~\cite{jiang2024motiongpt} and MoMask~\cite{guo2023momask} to generate corresponding human motions from text prompts. All motions are represented based on the SMPL~\cite{SMPL:2015} skeleton. 
Using the rendered human mesh normal map as a conditional input, we utilize the fine-tuned ControlNet~\cite{zhang2023adding} from HumanWild~\cite{ge20243d} to generate 2D HOI images.
For object templates, we utilize multiple approaches to obtain 3D objects from text prompts, including Text-to-3D models like Rodin~\cite{Rodin}, Meshy~\cite{wei2024meshlrm}, and LGM~\cite{tang2024lgm}, as well as online retrieval. The output object meshes and textures are directly used as the input object templates.
Since the generated 3D objects may not have metrically accurate scales, we adjust their scale to roughly align with the 2D HOI images. Fig.~\ref{fig:sameobj} illustrates our system's ability to handle objects from various sources with different scales.
As for physics simulation, our agents are trained on the IsaacGym~\cite{makoviychuk2021isaac} platform. We use the humanoid agent generated by \cite{Luo2022FromUH} based on the SMPL-X~\cite{cai2024smplerx} skeleton with a total actuated DoF of 51x3. Only the body skeleton of the reference SMPL are used as tracking rewards.
All training and inference is completed on a single RTX 4090 GPU. Specific prompts and parameter designs are presented in the supplementary document. 

\subsection{Zero-shot HOI Synthesis}
\label{subsec:zeroshot synthesis}
\subsubsection{Baseline Methods}
We compare our method against baseline methods CHOIS~\cite{li2023controllable} and HOI-Diff~\cite{peng2023hoi}. HOI-Diff synthesizes human-object interactions based on a text prompt and object geometry. CHOIS also generates human-object interactions but additionally requires sparse waypoints as input.
To ensure fairness, we used the results from our Part 1 as waypoints information required by CHOIS during comparison.

\subsubsection{Qualitative Results}
Given text prompts, our framework synthesizes human-object interaction results in a zero-shot manner.
We first showcase the final results using generated 2D images in Fig.~\ref{fig:results}.  To demonstrate the generalizability of our system, we have selected a variety of HOIs featuring objects of diverse shapes and distinct motion patterns, which are \textit{A man is playing the guitar. A man is lifting a barbell with both hands. A man is walking forward with an umbrella in his right hand. A man is holding a microphone in his right hand and speaking. A man gradually sits down on a chair.} In these examples, where there is significant variation in object shapes and human motion amplitudes, our system handles them well.

\begin{figure}[t]
    \centering
    \includegraphics[width=\linewidth]{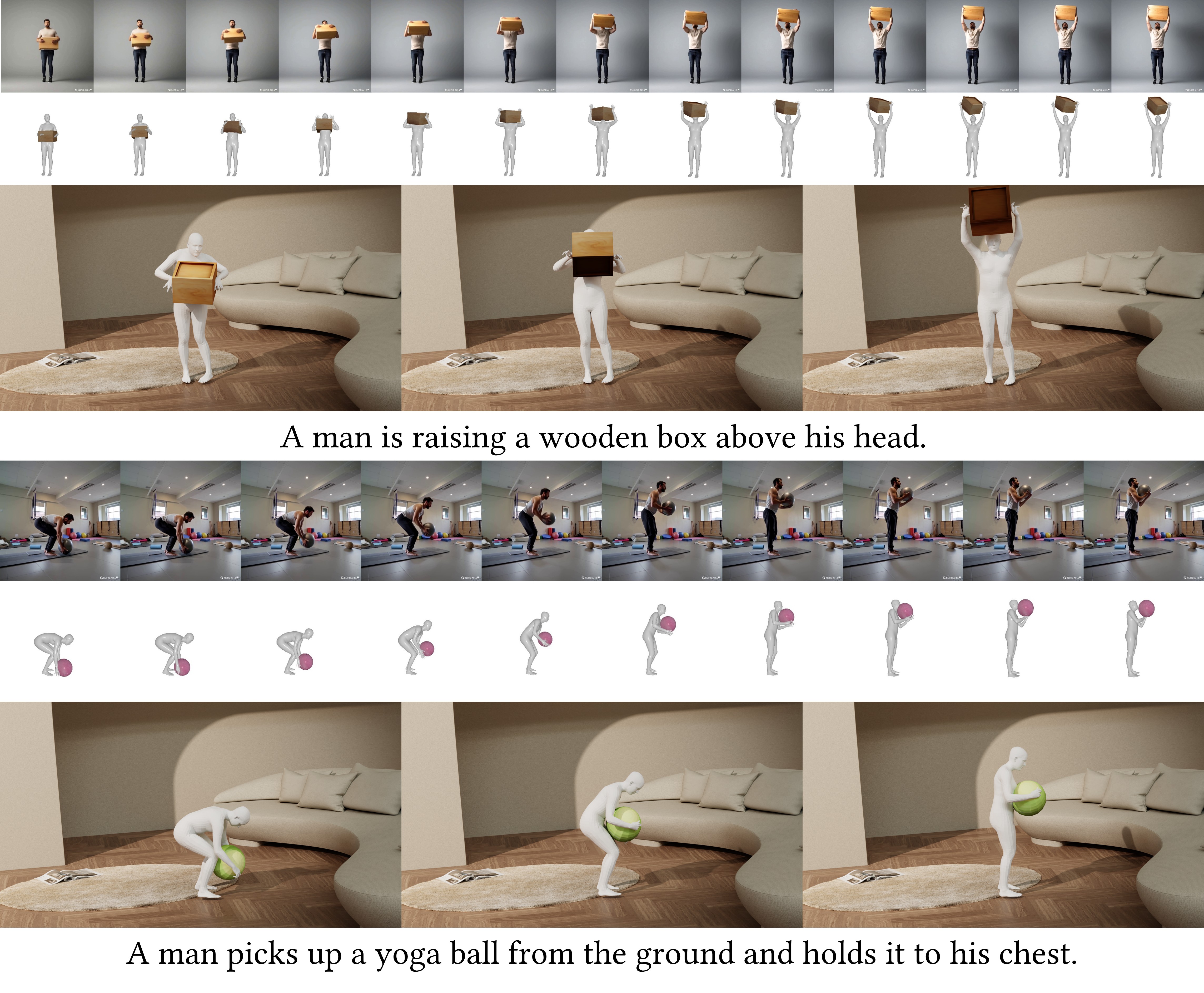}
    \caption{Zero-shot human-object interaction results generated by our system, using video generation models.}
    \Description{}
    \label{fig:video_gen_results}
\end{figure}

We then present the results of our system using generated 2D videos. In Fig.~\ref{fig:video_gen_results}, we showcase the generated 2D video, the intermediate 3D HOI milestones utilized as references for physics-based tracking, and the final results. 
In the first case of \textit{"A man is raising a wooden box above his head"}, we use the generated 2D image as an additional start-frame condition. This provides a better camera perspective for the system to estimate the object's pose, compared to the second case where only text prompts are used.
We refer readers to our Supplemental Video for a more detailed illustration of HOI motion quality.

\begin{figure}[t]
    \centering
\includegraphics[width=\linewidth]{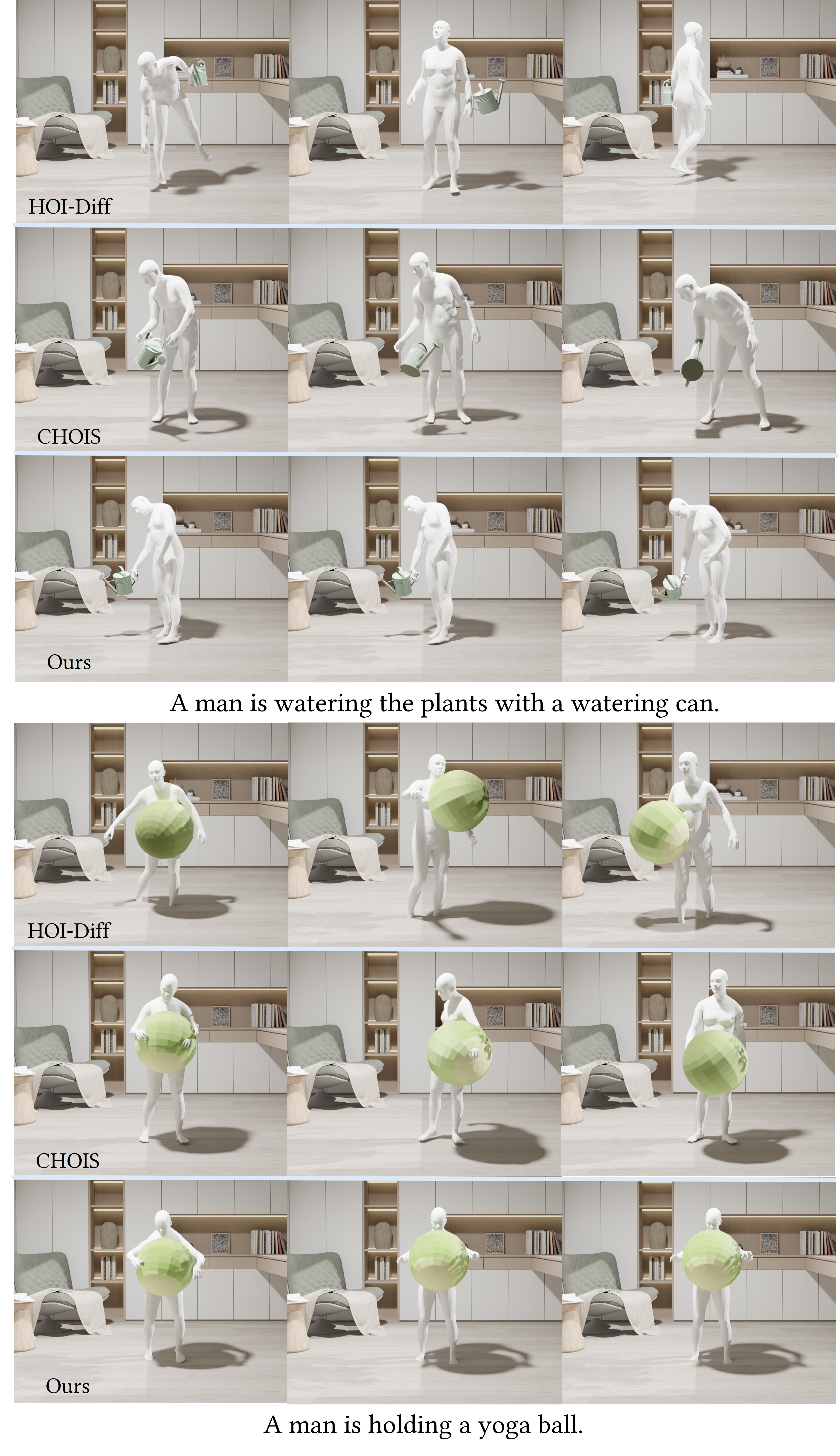}
    \caption{comparison results of our method against baseline methods CHOIS~\cite{li2023controllable} and HOI-Diff~\cite{peng2023hoi}. Please refer to the supplemental video for a clearer visualization of the HOI motion performance.
    }
    \label{fig:comparison_result}
\end{figure}

We also show comparison results of our method against CHOIS and HOI-Diff in Fig.~\ref{fig:comparison_result}.
Our method produces natural movements, while HOI-Diff and CHOIS fail to generate realistic or coherent interactions.

\subsubsection{Quantitative Results}
To evaluate the effectiveness of our method, we compare it with baseline works using various metrics and a user study. Quantitative results show that our approach significantly outperforms existing methods.

In our evaluation, we primarily focus on motion quality and adopt three key metrics: (1) \textbf{FS}: foot sliding score, which quantifies foot stability following~\cite{li2023controllable}; (2) \textbf{IV}: overlap volume between hand and object meshes, following~\cite{grady2021contactopt}; and (3) \textbf{CP}: contact percentage, which measures the proportion of frames where contact is detected, following ~\cite{li2023controllable}.

To provide a more intuitive evaluation of the final motion quality, we conducted a user study. We randomly selected 10 diverse text prompts and generated results using our method, CHOIS, and HOI-Diff. All results were rendered using our consistent rendering pipeline, presented together in randomized order, and evaluated by 20 users aged 16 to 30. Participants rated each result on a scale of 1 to 5 based on three criteria: alignment with the text prompt, physical realism, and overall quality. The final scores were averaged and summarized, as shown in Table~\ref{tab:user_study}.

Table comparing methods
\begin{table}[t]
\centering
\caption{Comparison of our method with HOI-Diff and CHOIS across quantitative metrics.}
\label{tab:metrics_comparison}
\resizebox{0.7\columnwidth}{!}{
\begin{tabular}{@{}lccc@{}}
\toprule
\textbf{Method}    & \textbf{FS}$\downarrow$ & \textbf{IV}$\downarrow$ & \textbf{CP}$\uparrow$ \\ \midrule 
\textbf{HOI-Diff}  & 3.50                 & 0.40                & 78.4                         \\ 
\textbf{CHOIS}     & 5.23                & 0.85               & 96.6                         \\
\textbf{Ours (w.o physics tracking)}      & 1.87                 & 0.35                & 89.3      \\
\textbf{Ours}      & \textbf{1.12}   & \textbf{0.15}          & \textbf{98.6}                        \\ \bottomrule
\end{tabular}
}
\end{table}

\begin{table}[t]
\centering
\caption{User study results comparing our method with HOI-Diff and CHOIS.}
\label{tab:user_study}
\resizebox{\columnwidth}{!}{%
\begin{tabular}{@{}lccc@{}}
\toprule
\textbf{Method}    & \textbf{Motion Quality} & \textbf{Physical Plausibility} & \textbf{Overall Rating} \\ \midrule
\textbf{HOI-Diff}  & 1.99                    & 1.38                        & 1.65                    \\ 
\textbf{CHOIS}     & 2.10                    & 1.57                          & 1.94                    \\ 
\textbf{Ours}      & \textbf{3.92}           & \textbf{4.31}                & \textbf{4.27}                    \\ \bottomrule
\end{tabular}%
}
\label{tab:user_study}
\end{table}

\subsection{System Analysis}
\label{subsec:system_analysis}
In this section, we analyze the effectiveness of the key components of our system, emphasizing the improvements introduced by our approach.

\subsubsection{Object Templates}
\begin{figure}[h]
    \centering
    \includegraphics[width=\linewidth]{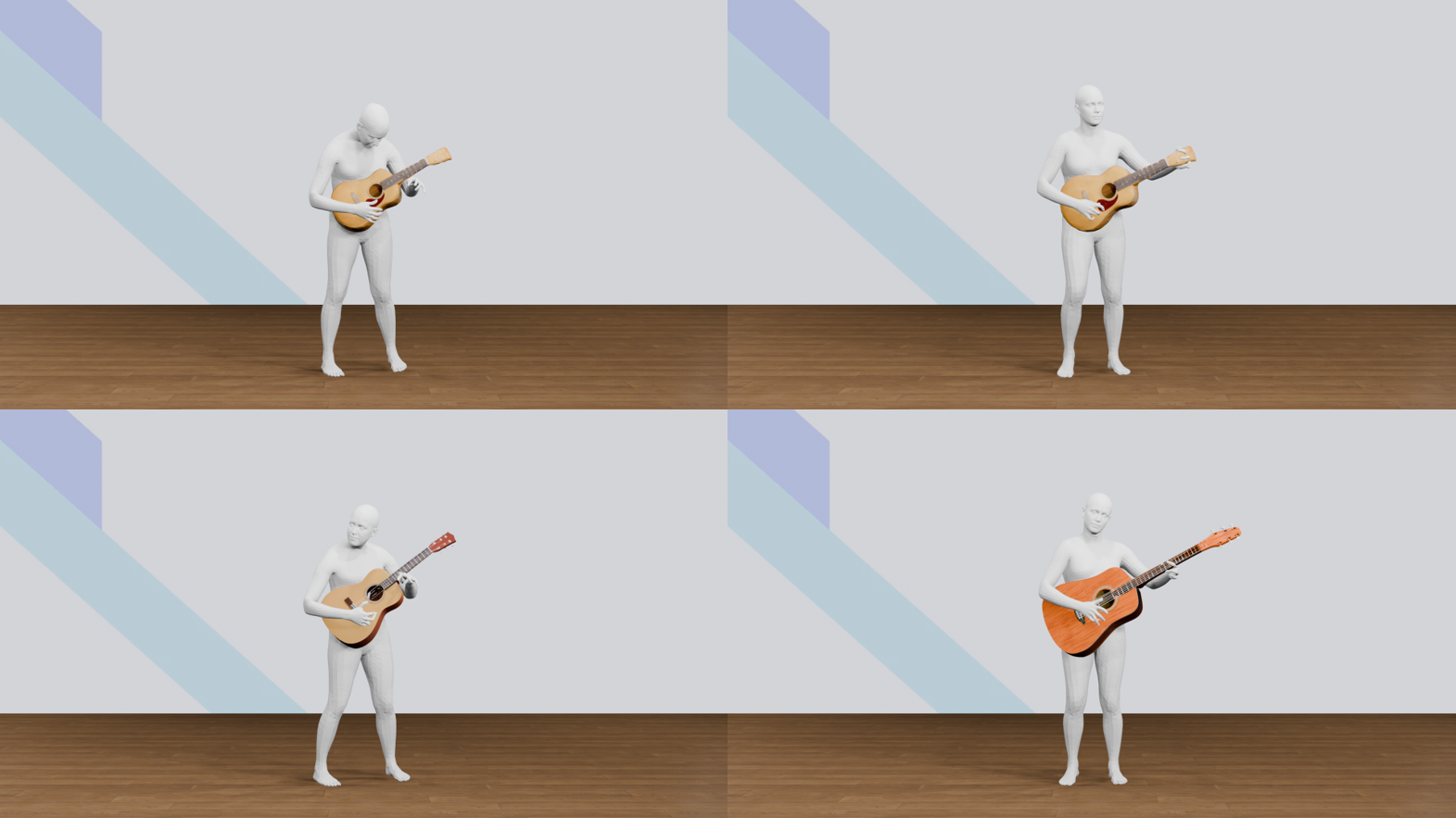}
    \caption{Given the same text prompts, our system supports the use of different object templates in various sizes.}
    \Description{}
    \label{fig:sameobj}
\end{figure}

Our system obtains the object templates from text prompt input employing Text-to-3D models or retrieving online, and shows adaptivity to object shape and appearance variance.  
As shown in Fig.~\ref{fig:sameobj}, given the same text prompt, we can use different objects obtained from the aforementioned sources or the same object in different sizes to produce plausible results.

\begin{figure}[h]
    \centering
    \includegraphics[width=\linewidth]{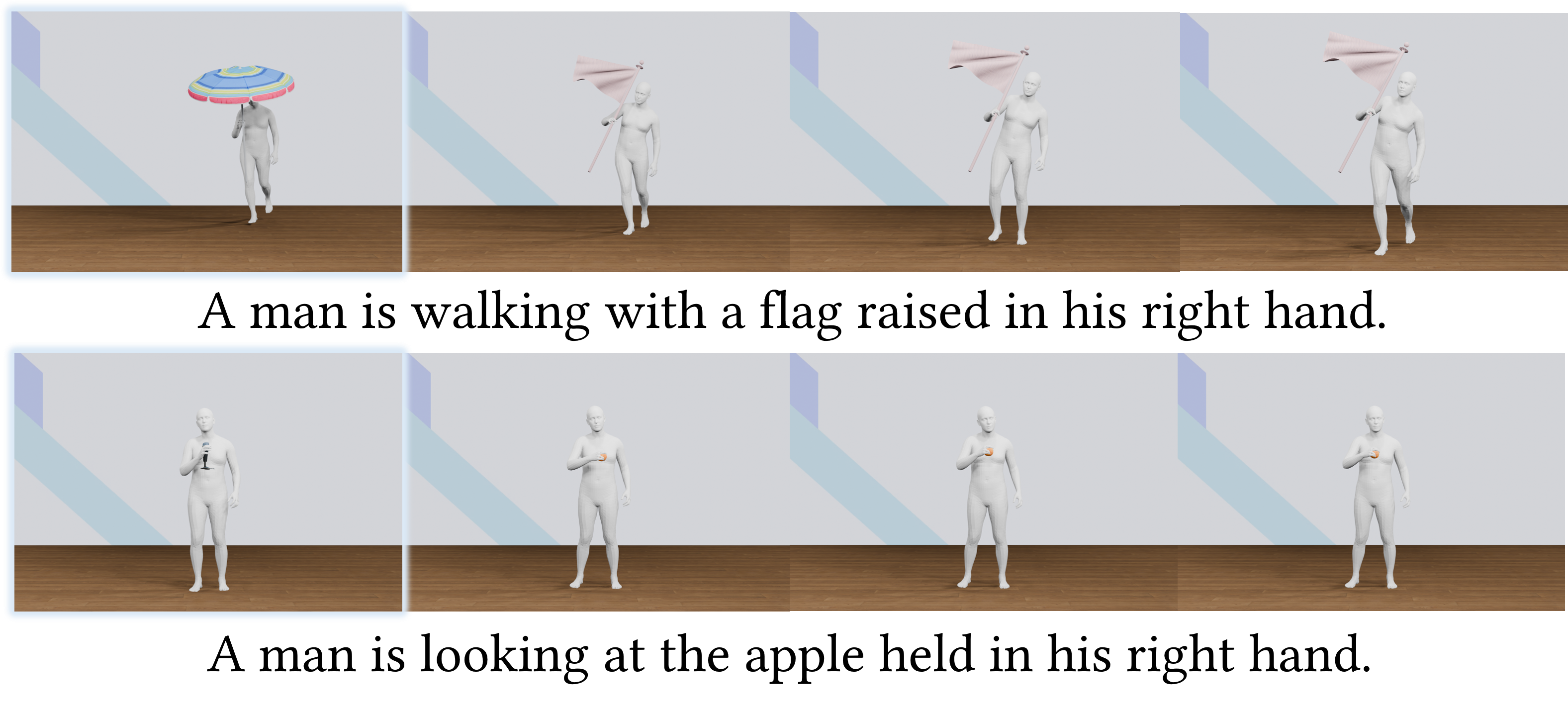}
    \caption{Using the same human motion as the image generation condition, our framework can produce HOI results that vary only in the object category. }
    \Description{}
    \label{fig:samemotion}
\end{figure}

Our generative 2D image pipeline enables controllable HOI generation by varying only the object while keeping the human motion fixed. For HOIs involving similar human motions and interaction patterns, the same human motion can be reused as the conditioning input for image generation. For example, in cases such as walking with an umbrella and walking with a flag, we use the same human motion generated from the prompt \textit{"A man is walking forward with an umbrella in his right hand"} and produce distinct HOI outcomes, as illustrated in Fig.~\ref{fig:samemotion}.
Note that human motion can also be sourced from diverse inputs, such as estimations from real-world videos. This capability also enables object-level editing in motion sequences with similar HOI patterns.

In the results of walking with an umbrella and a flag, we use the same kinematic motion sequence and change the prompt in the image generation phase to obtain new HOI results. 

\subsubsection{2D HOI Milestones}
We generate the 2D HOI milestone images using the obtained human pose from Text-to-Motion models as a condition (Sec.~\ref{subsubsec:gen_2d_image}). We here show a qualitative comparison with ComA~\cite{coma} which uses manually initialized object poses for different objects to estimate human poses. 
\begin{figure}[h]
    \centering
    \includegraphics[width=\linewidth]{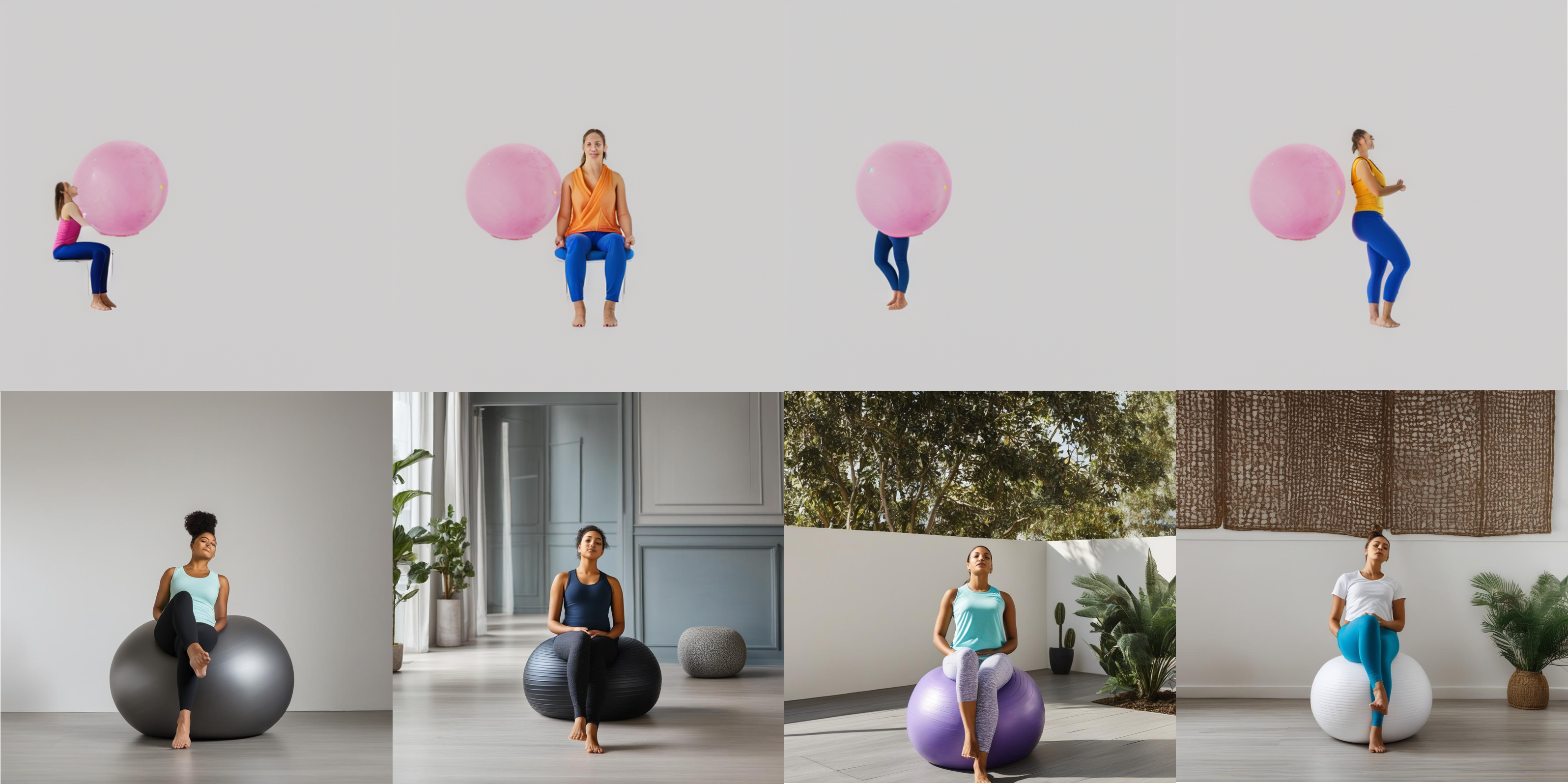}
    \caption{2D HOI Images from ComA (top) and Ours (bottom).}
    \Description{}
    \label{fig:coma2d}
\end{figure}
In Fig.~\ref{fig:coma2d}, we compare the results generated using the prompt "A person sits on a yoga ball" in our framework and ComA. We randomly selected four images from each for comparison. The results show that infering human pose based on the object produces low-quality images that do not align with the text prompt. This approach leads to information loss while it can generate common interaction patterns in 2D, it tends to collapse when dealing with less common text prompts. In addition, some objects can only produce meaningful inpainting results from specific angles. For example, an umbrella needs to appear with its canopy facing upward in the upper part of the image. However, defining the object's pose in such cases requires manual adjustments.

\subsubsection{Text-to-Motion Models}
As demonstrated in Sec.~\ref{subsec:human_motion_gen}, we apply a pre-trained pose estimation model to the generated 2D HOI images to extract the human pose, which serves as our final 3D HOI milestone. This step enhances existing Text-to-Motion models by leveraging 2D HOI priors to rectify the original object-agnostic pose.

\begin{figure}[h]
    \centering
    \includegraphics[width=\linewidth]{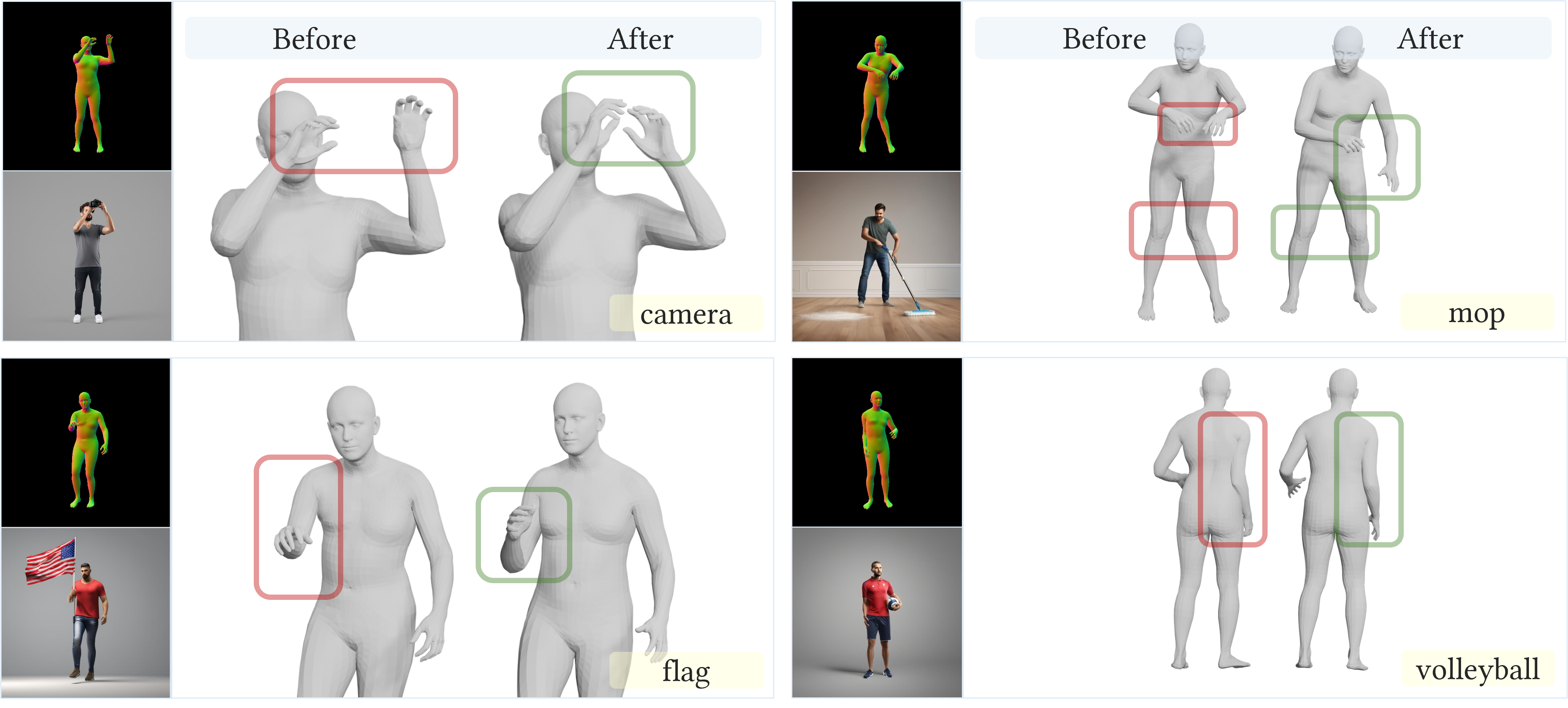}
    \caption{Our system rectifies the poses generated by Text-to-Motion models by leveraging 2D HOI priors.}
    \Description{}
    \label{fig:rec}
\end{figure}
We here show cases in Fig.~\ref{fig:rec} to demonstrate the effectiveness of our motion rectifying phase. For each case, we will display the input image, the generated result, and the corresponding human mesh for both.
In the case of the camera example, the generated motion has the hands positioned too far apart. If an object is added directly based on this motion, it becomes difficult for the camera to simultaneously contact both hands. However, in the generated images, the distance and orientation of the hands become more reasonable, and the new mesh derived from these images appears more realistic and natural. In the flag example, we can see that the pose and orientation of the right hand holding the flag have become more reasonable. As for the mop example, the original motion quality was quite poor. However, the newly obtained pose features hand movements that are more in line with the interactive nature, and there are no unnatural rotations in the knee joints. In the volleyball example, it can be noticed that the initial right arm was too tightly clamped, causing self-collision with the body. In contrast, the new arm position is more naturally aligned along the side of the body.

\subsubsection{3D HOI Milestones}
\begin{figure}[h]
    \centering
    \includegraphics[width=\linewidth]{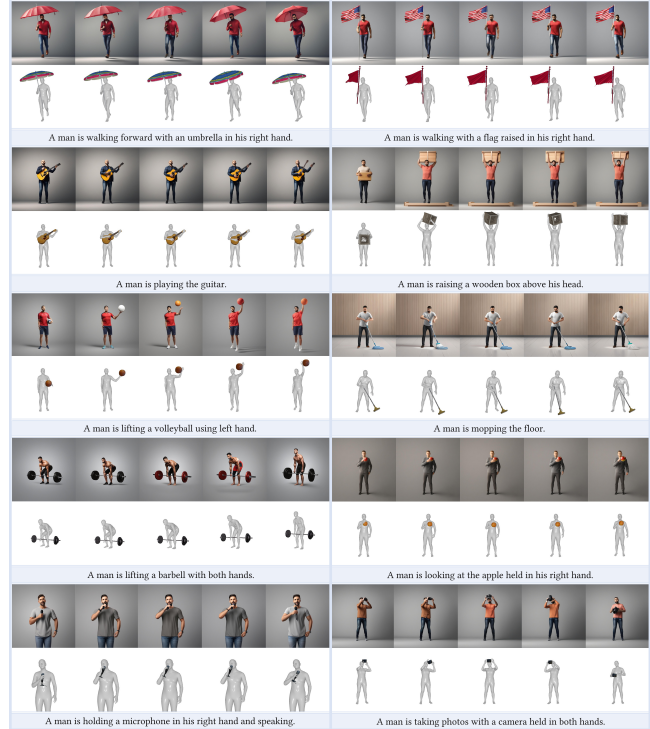}
    \caption{3D HOI milestone results from our system, using image generation models.}
    \Description{}
    \label{fig:part1_results}
\end{figure}

The quality of the final human-object interaction results largely depends on the quality of our 3D HOI milestones of human and object poses. We show the results in Fig.~\ref{fig:part1_results} obtained from 2D generative image pipeline. The results from 2D generative video pipeline can be found in Fig.~\ref{fig:video_gen_results}.

\begin{figure}[h]
    \centering
    \includegraphics[width=\linewidth]{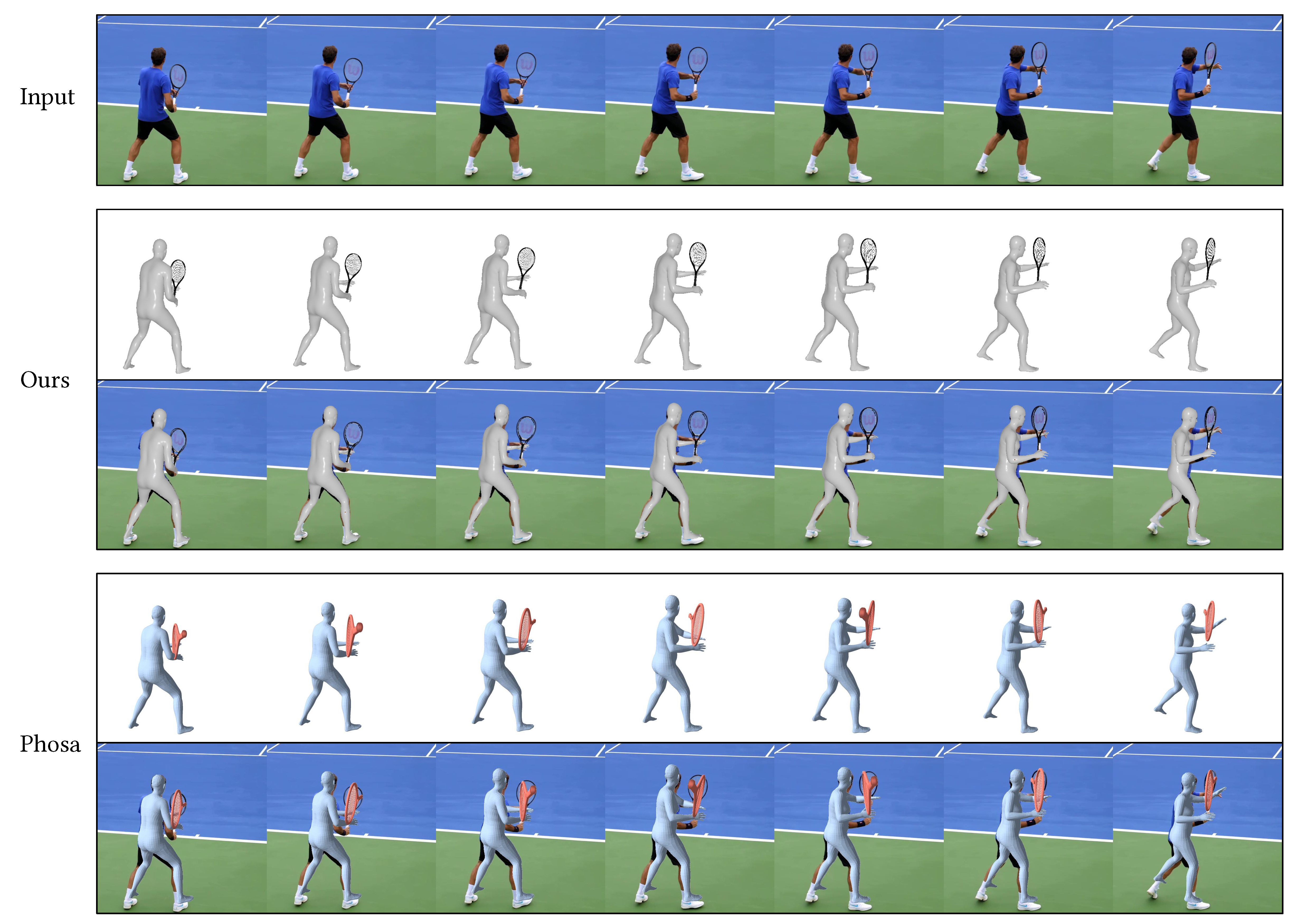}
    \caption{Object 6-DoF pose estimation comparison with PHOSA.}
    \Description{}
    \label{fig:comparison_phosa}
\end{figure}

Although our generalizable category-level object 6-DoF estimation method is tailored for our system to work with 2D HOI images, it can also be applied to estimate object poses in real-world images. To assess its performance, we compared it with another optimization-based method PHOSA~\cite{zhang2020perceiving} on an in-the-wild video. As illustrated in Fig.~\ref{fig:comparison_phosa}, our method achieves more accurate pose estimation.  This improvement is mainly attributed to our framework's integration of coarse pose estimation which leverages semantic correspondence from a pre-trained 2D vision model~\cite{oquab2023dinov2}, and differentiable rendering utilizing a pre-trained depth estimation method~\cite{yang2024depth}.

\subsection{Failure Cases}
\label{subsec:failure}
Our system is designed with the capacity to mitigate errors at one stage through subsequent corrections in later stages.
For example, the human poses from Text-to-Motion models will be rectified using 2D HOI images, and the physics-based tracking module helps mitigate the inaccuracies in 3D HOI milestones.

The performance of our system is highly contingent on the quality of the multi-modal outputs generated from text prompts.
In practice, we typically sample approximately three generations from the multi-modal models to obtain satisfactory results.
In the following discussion, we will present examples of failure cases resulting from unsatisfactory multimodal model outputs and highlight scenarios that exceed the capabilities of our system.

First, for the outputs of Text-to-3D models, a watertight mesh is required, particularly for the physics-based tracking component. A non-watertight mesh results in incorrect collision calculations within the physics environment, leading to undesired outcomes.

Second, for the generated 2D HOI milestones, it is crucial that the camera view ensures the object appears sufficiently large to provide enough information for accurate pose estimation. Objects that are too small or lack distinct texture information lead to failed pose estimation.
Additionally, the generated HOI should be reasonable and aligned with realistic interactions. However, we occasionally observe unrealistic results from video or image generation models, such as playing tennis with both hands or exhibiting overly discontinuous object and human movements.

Besides the scenarios that lead to failure cases due to multimodal output, there are specific HOI cases that our model struggles to handle effectively. One such case involves discontinuous contact, such as playing basketball, where the interaction includes intermittent contact between the human and the object. Another challenge lies in complex object manipulations, such as tying shoelaces or assembling small parts, which require precise modeling of hand-object interactions beyond the current capabilities of our framework. These limitations underscore areas for future development and enhancement.



\section{Conclusion}
\label{sec:conclusion}
In this paper, we propose a novel zero-shot method for generating human-object interactions (HOIs) without relying on 3D HOI datasets, addressing the limitations of existing methods in terms of object diversity and interaction patterns. Our system leverages existing HOI priors from pre-trained multimodal models to generate coarse 3D HOI kinematic motion. By refining this motion with a physics-based tracking strategy, our approach produces open-vocabulary HOIs with enhanced physical realism.
The results demonstrate the potential of our method for scalable and diverse HOI generation.  

There is still room for improvement in our current research.  
Firstly, the performance and success rate of our pipeline are significantly constrained by the quality of the generated HOI priors, such as the 2D images and videos. Poor initial image generation can lead to degraded performance in subsequent stages.
Secondly, our system does not explicitly model detailed hand movements due to their complexity, which limits its ability to handle intricate object manipulations. A potential solution to overcome this limitation is to adopt the GAIL framework to train a hand-specific discriminator, allowing for more precise handling of hand interactions and significantly enhancing the realism of the generated HOIs.

\bibliographystyle{ACM-Reference-Format}
\bibliography{hot}

\clearpage
\setcounter{page}{1}

\section{Text Prompts}

\subsection{Text Prompt for 2D Generation}
We use the text prompts as input to Text-to-3D, Text-to-Motion, Text-to-Image, Text-to-Video models. The primary text prompt for each case consists of a simple description of the human-object interaction, such as \textit{"a man is holding a yoga ball"}. 
This simple text prompt serves as the input for Text-to-Motion models, while the object description (e.g., \textit{"yoga ball"}) within the prompt is used as input for Text-to-3D models.
Auxiliary prompts include parameters like \textit{best quality}, \textit{realistic} and \textit{simple background} are provided as input to the 2D diffusion models. For the Text-to-Video models, we include additional prompts to control the camera settings, such as \textit{"use a constant camera view, without zooming in or out. The camera captures the whole body of the person from the side."} 
This textual description of the camera is unnecessary for Text-to-Image models, as they rely on rendered human normal maps with predefined camera settings as input. This advantage in image generation also enhances the video generation results that utilize an additional start-frame image condition for guidance.

\subsection{Text Prompt for LLM}
As mentioned earlier, we utilize LLMs to acquire contact information between humans and objects. Specifically, we primarily use GPT-4o~\cite{gpt4o} and LLAMA~\cite{dubey2024llama} for this phase.
We find LLMs struggle to provide precise, time-sequential contact information. Therefore, we only use LLms to determine which body parts remain in constant contact with the object and which never make contact throughout the motion. Our system utilize SMPL-X~\cite{cai2024smplerx} which includes 51 joints for our human model. For these joints, we assign \textit{contact} and \textit{separate} labels to compute the contact reward. 
We now provide a detailed explanation of the prompts used to obtain the contact labels.

For a given motion \( X \), we design prompts as follows:  
\textit{In motion \( X \), involving an object \( Y \), which body parts remain in constant contact with \( Y \), and which body parts never make contact (especially those prone to accidental collisions)? Please classify only from the following body parts: Pelvis, L\_Knee, L\_Ankle, L\_Toe, R\_Knee, R\_Ankle, R\_Toe, Torso, Chest, Neck, Head, L\_Shoulder, L\_Elbow, L\_Wrist, R\_Shoulder, R\_Elbow, and R\_Wrist. No additional description is required. Respond strictly in the following format: contact:["L\_Wrist"], separate:["R\_Elbow"].}

For large models that support visual input, uploading images corresponding to the motion can further enhance the accuracy of the results. This approach simplifies the labeling process while ensuring relevant contact information is captured for reward computation.

When estimating the object pose, we also use contact information. Since we only consider whether the hands are in contact with the object, the prompt is as follows:  
\textit{In motion \( X \), involving an object \( Y \), does the person’s left hand or right hand remain in constant contact with the object? Provide a True or False judgment in the format: “Left Hand: True, Right Hand: False”}.

\section{Zero-shot HOI Generation}

\subsection{2D HOI Images Generation}
When rendering human mesh normal map, we standardize the depth of the human mesh's root joint from the camera as well as the horizontal displacement.
This approach ensures that when the same seed value is used, the generated images within a sequence are more consistent in terms of the positioning and orientation of the human mesh relative to the camera. 
For camera orientation, we align it based on the rotation in the first frame, ensuring that the human body is facing the camera in the initial frame.
We use Humanwild\cite{ge2024humanwild} to generate 2D HOI images conditioning on the rendered human normal map. Due to the unstable quality of our input images, we appropriately reduced the conditioning scale from the default 0.5 to 0.3 to achieve more natural human poses and better facilitate object completion.

\subsection{Object 6-DoF Pose Estimation}
\begin{figure}[h]
    \centering
    \includegraphics[width=\linewidth]{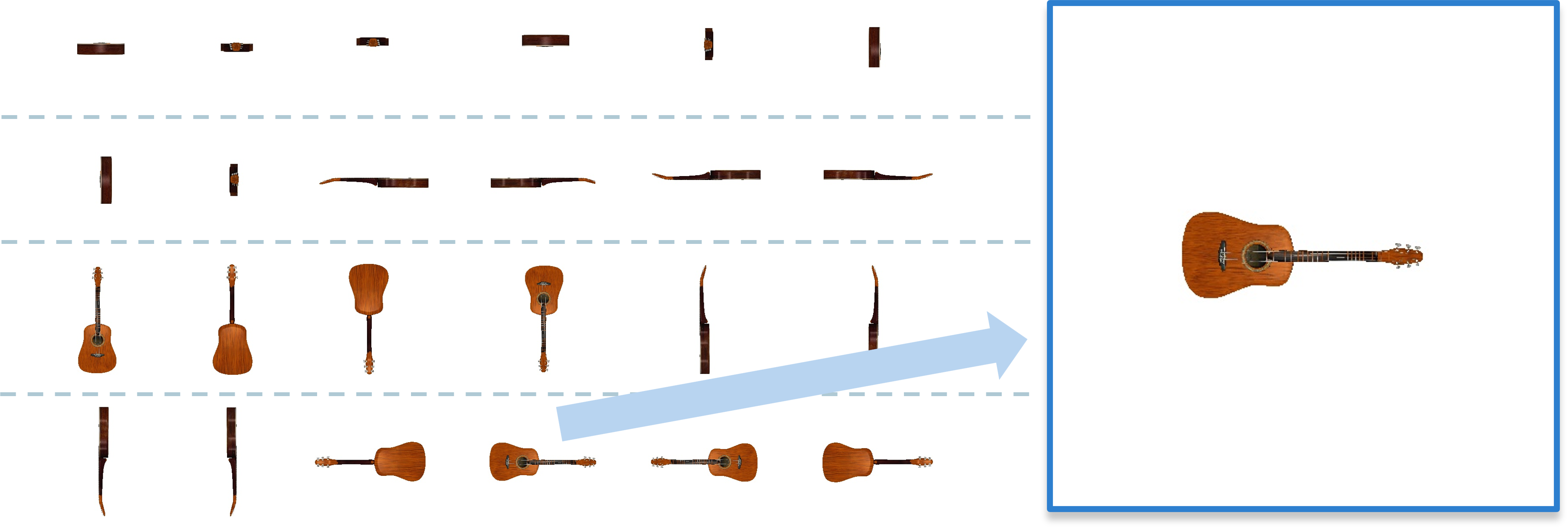}
    \caption{Initialize Selector. We renders the object from 24 different viewpoints and then select the viewpoint with the highest similarity between the rendered object image and the 2D HOI image.}
    \Description{}
    \label{fig:selector}
\end{figure}

We utilize the Pytorch3D~\cite{ravi2020pytorch3d} differentiable renderer to implement our object pose estimation method. The rendered image resolution is set to 
512 x 512, with a camera focal length of 700 and the principal point located at the image center. The camera is positioned at the origin with an identity rotation matrix.
\paragraph{Initial View Selector}
As shown in Fig.~\ref{fig:selector}, our viewpoint selector employs a systematic strategy to ensure comprehensive coverage of the entire object.
With the camera's position at the origin and the object positioned at the center of the human mesh, we select 24 specific rotations derived from the symmetry group of the cube, known as the octahedral group (\(O_h\)), a subgroup of the full 3D rotation group (\(SO(3)\)). We use the Efficient Perspective-n-Point~\cite{epnp} algorithm implemented in Pytorch3D and the Plane Homography algorithm with Ransac in Opencv~\cite{opencv}.

\paragraph{Differentiable Rendering}
To achieve more stable optimization, we adopt a multi-stage optimization approach using the proposed losses described in Sec~\ref{subsubsec:diff_rendering}. In the first stage, we optimize only the object and human-object silhouette losses to correct alignment errors resulting from the coarse pose estimation using semantic correspondence and Perspective-n-Point (PnP) algorithm.  
In the second stage, we incorporate the depth losses alongside the silhouette loss to refine the object's depth. In the final stage, we utilize all the losses including the human-object interaction losses to achieve joint optimization.
We use the Adam optimizer~\cite{kingma2014adam} with a learning rate of $1\times10^{-3}$ throughout training. Each stage is optimized for 200 iterations, with the total optimization for a single frame taking approximately 5 minutes on a single NVIDIA 4090 GPU.
The corresponding loss weights $w_i$ for each loss term $(L_i)$ as discussed in Sec~\ref{subsubsec:diff_rendering} and Sec~\ref{subsubsec:hoi_optimization} are specified as follows: $w_{sil}=100, w_{depth}^{rel} = 0.5, w_{depth}^{abs}=0.1, w_{contact}=1, w_{penetration}=100$, and $\theta = 0.1$, where $w_{sil}$ is the weight for the silhouette loss, $w_{depth}^{rel}$ and $w_{depth}^{abs}$ are the weights for the relative depth loss and metric human depth loss, $w_{contact}$ is the contact loss weight, $w_{penetration}$ is the weight for penalizing penetration, and $\theta$ is a predefined threshold for valid contact regions.

\section{Physics}

\begin{table}[h!]
\centering
\begin{tabular}{@{}lll@{}}
\toprule
\textbf{Parameter}          & \textbf{Description}                     & \textbf{Value} \\ \midrule
solver\_type               & Solver type             & TGS    \\
num\_position\_iterations  & Number of position iterations            & 4    \\
contact\_offset            & Contact offset                           & 0.01 \\
gravity                    & Gravity vector (m/s\textsuperscript{2})  & (0, -9.81, 0) \\
staticFriction             & Static friction coefficient              & 1.0  \\
dynamicFriction            & Dynamic friction coefficient             & 1.0  \\
restitution                & Restitution coefficient                  & 0.0  \\
density                    & Object density (kg/m\textsuperscript{3}) & 100.0 \\ 
\bottomrule
\end{tabular}
\caption{Simulation Environment Parameters}
\label{tab:simulation_parameters}
\end{table}

We follow the actor-critic framework widely used in previous work~\cite{peng2021amp}. The policy output is modeled as a Gaussian distribution of dimensions $51 \times 3$ with constant variance, and the mean is modeled by a two-layer MLP of [1024, 512] units and ReLU activations. The action at $\mathbb{R}^{51\times3}$ sampled from the policy is the target joint rotations for the PD controller. The PD controller adjusts and outputs the joint torques to reach the target rotations. The observation of our agent includes the rotation and position of the root, the rotation and angular velocity of all joints, the position and linear velocity of selected key joints, as well as the reference for these variables. Additionally, it contains the motion information of the object and the target pose.
Its GPU acceleration can simultaneously train agents in 4096 environments. For each action, convergence takes approximately 1 to 2 hours on a single RTX 4090 GPU, depending on the complexity of the motion.

The simulation and PD controller run at 60 Hz, with the policy sampled at 30 Hz. Humanoids and objects are initialized at the start using fixed rotations and root positions from the first reference frame. Random initialization is avoided to prevent severe collisions in HOI data that may eject objects. Early termination is enabled by kinematic state errors. As Isaac Gym lacks collision detection, contact situations are inferred from forces, which may occasionally introduce errors.

\begin{table}[h!]
\centering
\caption{Reinforcement Learning Parameters and Hyperparameters}
\label{tab:rl_parameters}
\begin{tabular}{@{}p{0.32\linewidth}p{0.12\linewidth}|p{0.32\linewidth}p{0.12\linewidth}@{}}
\toprule
\textbf{Parameter}          & \textbf{Value}          & \textbf{Parameter}         & \textbf{Value}         \\ \midrule
Learning Rate               & $2 \times 10^{-5}$       & Discount Factor ($\gamma$) & 0.99                   \\
Entropy Coefficient        & 0.01                     & Clip Range                 & 0.2                    \\
Termination Distance        & 0.50                    & Termination Height         & 0.30                    \\
$\lambda_{reg},\lambda_{acc}$ & -0.01          & $\lambda_{contact}$ & -3.0                    \\
$\lambda_p$ (position) & -1.0                     & $\lambda_r$ (rotation) & -0.3                   \\
$\lambda_v$ (linear vel.)   & -0.02                     & $\lambda_\omega$ (ang. vel.) & -0.02                   \\ \bottomrule
\end{tabular}
\end{table}

Our method uses convex hull decomposition to simplify collision handling by limiting the maximum number of convex shapes to 12. This approach balances computational efficiency with collision accuracy, ensuring realistic and efficient simulations.

For detailed simulation parameters and training parameters, please refer to Table~\ref{tab:simulation_parameters} and Table~\ref{tab:rl_parameters}.

\end{document}